\documentstyle[12pt]{article}
\begin{document}
\begin{center}
{\large\bf GRAVITATIONAL FIELD AND EQUATIONS OF MOTION
                    OF NONLINEAR COSMIC STRING}
\vspace{\baselineskip}

{\bf L.M.Chechin}$^1$
and
{\bf T.B.Omarov}$^2$
\end{center}

\begin{center}
Astrophysics Institute, Academy of Sciences of Kazakstan\\
Almaty 480068 Kazakstan\\
{\small $^1$ chel@afi.academ.alma-ata.su},
{\small $^2$ omar@afi.academ.alma-ata.su}
\end{center}
\vspace{2\baselineskip}
\begin{center}
August 30, 1998
\end{center}
\vspace{2\baselineskip}

\begin{center}
{\bf Abstract}
\end{center}
\begin{quotation}
{For the case of tension tensor containing nonlinear terms in
$l^{\alpha}$, we give generalization of Vilenkin metrics and
equations of motion of cosmic string. Dynamics of nonlinear
string in (1+1)-dimensional universe is discussed.}
\end{quotation}
\vspace{\baselineskip}
PACS: 98.80.Cq, 04.25.-g

\newpage
\section{Introduction}

Consider a stringful medium meaning a variety of thin, infinite
non-intersecting strings which fill up all given space-time $V_4$
in a continuous way. Energy-momentum tensor of the stringlike
medium is of the form [1,2]
\begin{equation}
T^{\alpha\beta} = \mu(u^{\alpha}u^{\beta} - l^{\alpha}l^{\beta}),
\end{equation}
where $u^{\alpha} = \displaystyle\frac{dx^{\alpha}}{d\tau}$
is time-like $(u^{\alpha}
u_{\alpha} < 0)$, and
$l^{\alpha} = \displaystyle\frac{dx^{\alpha}}{d\rho}$ is space-like
$(l^{\alpha}l_{\alpha} > 0)$ vectors. Vector $u^{\alpha}$ describes
dynamics of a separate string of the medium while $l^{\alpha}$
describes its space orientation.

We rewrite the tensor (1) in the following form:
\begin{equation}
T^{\alpha\beta} = \mu u^{\alpha}u^{\beta} - t^{\alpha\beta},
\end{equation}
where
\begin{equation}
t^{\alpha\beta} = \mu l^{\alpha}l^{\beta},
\end{equation}
and consider (3) as a tension tensor of the stringful medium arising
as the result of oscillation of separate string. Indeed, from
conservation of the energy-momentum
$\nabla_{\beta}T^{\alpha\beta} = 0$, taking into account
continuousness condition of the medium and each string separately,
we have equations of motion of a separate string
\begin{equation}
\frac{D^2x^{\alpha}}{d\tau^2} - \frac{D^2x^{\alpha}}{d\rho^2} = 0.
\end{equation}
Solutions of these equations, with initial or initial and boundary
conditions, are, as it is wellknown, travelling and standing
waves, respectively.
Therefore, these are such waves that create abovementioned tensions
in the medium.
In the case, considered oscillations are not only small but also
linear. Thus, expression (3) is a linear tension tensor of the
stringful medium.

In paper [3], using energy-momentum tensor (2) with a linear
tension tensor, the metrics of gravitational field created by
straight cosmic string has been obtained.
In subsequent papers, metrics of $V_4$ created by, for example,
rotating cosmic strings [4], cosmic strings with propagating kinks
along it [5], oscillating strings in the form of standing
waves [6] were considered.
In these papers behavior of light rays and test particles was
investigated [7-10].
In addition, equations of motion of cosmic strings (4) in some
external [11, 12] and selfconsistent [13, 14] gravitational fields had
been studied. However, all these results, being obtained on the basis
of equations (1)-(4), have a linear character too.

We remind, however, that cosmic strings are supposed to appear
at early stages of the evolution of universe from the scalar fields
arisen as the result of phase transitions in vacuum [15].
Simplest lagrangian describing such fields is the nonlinear Higgs
lagrangian for complex scalar field $\chi(x)$,
\begin{equation}
{\cal L} = \partial_{\alpha}\chi^{\ast}\partial^{\alpha}\chi
+ m_0^2 \chi^{\ast}
\chi - \lambda(\chi^{\ast}\chi)^2 .
\end{equation}
So, nonlinear character of the field ($\lambda$ is a self-coupling
constant) should lead to nonlinearity of the cosmic string
generated by this field. Particularly, oscillations of such
a string should be essentially nonlinear.
It follows that the tension tensor of the stringful medium will have
nonlinear contributions. Let's construct nonlinear tension tensor
of the stringful matter.
\vspace{\baselineskip}
\section{Nonlinear tension tensor}

Consider a single oscillating string in the medium at some fixed
time, i.e. a position of the string on hypersurface $\tau = const$.
This hypersurface is parametrised by the variable
$\rho$ (see Fig.1). In Fig.1, all coordinate axises are shown,
for simplicity, as a single line.

Let some point $A$ on the string, in an equilibrium state, is given
by coordinates $X_{o}^{\alpha}$.
Under oscillations, this point takes position $X^{\alpha}$.
Denote the deflection of point $A$ from the origin by
$x_{1}^{\alpha}$, i.e.
\begin{equation}
x_{1}^{\alpha} = X^{\alpha} - X_{o}^{\alpha}.
\end{equation}
Supposing that oscillations are small, we have
\begin{equation}
\frac{x_{1}^{\alpha}}{X^{\alpha}} \simeq
\frac{x_{1}}{X_{o}^{\alpha}} = \varepsilon
\ll 1.
\end{equation}
Therefore,
\begin{equation}
x_{1}^{\alpha} = \varepsilon X^{\alpha},
\end{equation}
where $\varepsilon$ is a small parameter.
It can be introduced, for example, as a ratio between
oscillation amplitude $a$ of string and half-length of wave $L/2$,
i.e.$\varepsilon = 2a/L$. Using definition of
$l^{\alpha}$, one can show that deflection of oscillating string
from the equilibrium state, due to (8), is described by the vector
\begin{equation}
l_{1}^{\alpha}
= \frac{dx_{1}^{\alpha}}{d\rho}
= \varepsilon\frac{dX^{\alpha}}
{d\rho} = \varepsilon l^{\alpha},
\end{equation}
where  $l^{\alpha}$
is the vector defining orientation of the string.
Since the string is one-dimensional object, its oscillations
produce tension along the string. So, the tension vector
$S^{\alpha}$ is proportional to $l^{\alpha}$, i.e.
\begin{equation}
S^{\alpha} = \stackrel{\ast}{\mu}l^{\alpha},
\end{equation}
where $\stackrel{\ast}{\mu}$ is some coefficient.
Oscillations along each of coordinate axises $x^{\alpha}$
are due to projections of the tension vector
$S^{\alpha}$ on the corresponding axis,
i.e. due to its normal component $S_{(n)}^{\alpha}$.
Its components on coordinate axises are components of
tension tensor $t^{\alpha\beta}$. From Fig.1 and (10)
it can be seen that
\begin{equation}
t^{\alpha\beta}
= S^{\alpha}\sin\psi^{\beta}
= \stackrel{\ast}{\mu}l^{\alpha}
\sin\psi^{\beta},
\end{equation}
where angle $\psi^{\beta} = (\widehat{x^{\beta};l^{\alpha}})$.
Again, from Fig.1 it follows that
\begin{equation}
\tan{\psi^{\beta}} = \frac{dx_{1}^{\beta}}{d\rho} = l_1^{\beta}.
\end{equation}
For small angles the following approximation is valid
(up to third power of the angle):
$$ \sin\psi^{\beta}
= \frac{\tan\psi^{\beta}}{\sqrt{1 + \tan^{2}\psi^{\gamma}}}
\simeq \tan\psi^{\beta}(1 - \frac{1}{2}\tan^{2}\psi^{\gamma}),$$
which, in accordance to (12),  can be written as
\begin{equation}
\sin\psi^{\beta} \simeq l_{1}^{\beta}(1 - \frac{1}{2}l_{1}^{\gamma}
l_{1}^{\gamma}).
\end{equation}
Inserting this equation to (11), using the earlier found
relation (9), and restricting consideration by the
abovementioned approximation we obtain the nonlinear tension tensor:
\begin{equation}
t^{\alpha\beta}
= \mu l^{\alpha}l^{\beta}(1 - \frac{\varepsilon^2}{2}l^{\gamma}
l_{\gamma}),
\end{equation}
where $\mu = \varepsilon\stackrel{\ast}{\mu}$
is linear mass  density of the strings.
\vspace{\baselineskip}
%
\section{Gravitational field of nonlinear cosmic string}

 From eqs. (2) and (14) one can derive nonlinear energy-momentum
tensor of the stringful medium, namely,
\begin{equation}
T^{\alpha\beta} = \mu[u^{\alpha}u^{\beta} - l^{\alpha}l^{\beta}(1 -
\frac{\varepsilon^2}{2}l^{\gamma}l_{\gamma})]
\end{equation}
and use it to find metrics of gravitational field created by
alone nonlinear cosmic string.

All further calculations will be made with supposing that
gravitational field is weak, i.e., as usually, taking
$g_{\alpha\beta} = \delta_{\alpha\beta} +
h_{\alpha\beta}$, where $h_{\alpha\beta}$
are small contributions to pseudo-Euclidean metrics
$\delta_{\alpha\beta}$. Then, Einstein equations
can be linearised, and in harmonic coordinates take the wellknown
form
\begin{equation}
\Box h_{\alpha\beta} = -16\pi\gamma{\cal T}_{\alpha\beta},
\end{equation}
where ${\cal T}_{\alpha\beta}
= T_{\alpha\beta} - \frac{1}{2}\delta_{\alpha\beta}T$.
To derive ${\cal T}_{\alpha\beta}$ one have to take into account
that vectors $u^{\alpha}$  and  $l^{\alpha}$
obey the following orthonormal gauge conditions:
\begin{equation}
u^{\alpha}u_{\alpha} + l^{\alpha}l_{\alpha} = 0,
u^{\alpha}l_{\alpha} = 0.
\end{equation}
Thus, taking into account (17) we have finally
\begin{equation}
{\cal T}^{\alpha\beta}
= \mu[u^{\alpha}u^{\beta} - l^{\alpha}l^{\beta}
(1 - \frac{\varepsilon^2}{2})
+ \delta^{\alpha\beta}l^{\lambda}l_{\lambda}
(1 - \frac{\varepsilon^2}{4}
l^{\gamma}l_{\gamma})]\delta_{3}(x - x_0).
\end{equation}

Now, we put additional restrictions on the form of space-time
interval under study. Namely, we will find it in the orthogonal
form. For this purpose, we take $l^0 = 0$.
Besides, we will suppose that cosmic string is immobile as whole,
i.e. $u^k = 0$. Under these conditions, equations (16)
in the component form are:
\begin{equation}
\Box h_{00} = 4\pi\gamma\varepsilon^2\delta_{00}\delta_{3}(x - x_0),
\end{equation}
\begin{equation}
\Box h_{kl} = 16\pi\gamma\mu[l_kl_l(1
+ \frac{\varepsilon^2}{2}l^ml_m) +
\delta_{kl}l^nl_n(1
+ \frac{\varepsilon^2}{4}l^ml_m)]\delta_{3}(x - x_0).
\end{equation}

Equations (19)-(20) have the form of D'Alambert equations, so
its solutions are retarded potentials. Expanding the potentials
in power series of the parameter
$\bigl|x - x'\bigr|/x^0$ and retaining only leading terms
we find timelike component,
\begin{equation}
h_{00}
= - \gamma\mu\varepsilon^2\delta_{00}\int\limits^{}_V\frac{\delta_3
(x - x')}{\bigl|x - x '\bigr|}dV'
= -2\gamma\mu\varepsilon^2\ln(\frac{r}{r_0}).
\end{equation}

To find spacelike components we need to fix orientation of the
string. We suppose that the string is placed along the
$Oz$ axis. Then,
$l^1 = l^2 = 0, l^3 = 1$ and so
$$h_{kl} = - 4\gamma\mu\int\limits^{}_{V}[{l_k}'{l_l}'(1
+ \frac{\varepsilon^2}{2}
{l'^m}{l'_m}) + $$
\begin{equation}
\delta_{kl}{l'}^n{l'}_n(1
+ \frac{\varepsilon^2}{4}{l'}^m{l'}_m)]\bigl|x - x'\bigr|^{-1}
\delta_3(x - x_0)dV' =
 -8\gamma\mu(1
+ \frac{\varepsilon^2}{4})\ln(\frac{r}{r_0})\delta_{kl}.
\end{equation}

Subsequently, the interval of gravitational field created by
nonlinear cosmic string has the form:
\begin{equation}
dS^2 = (1 - 2\gamma\mu\varepsilon^2\ln(\frac{r}{r_0}))d{x^0}^2 -
(1 + 8\gamma\mu(1 + \frac{\varepsilon^2}{4})
\ln(\frac{r}{r_0}))\delta_{kl}
dx^kdx^l.
\end{equation}
At the limit $\varepsilon \to 0$, the eq.(23)
reproduce the Vilenkin's interval [3].
\vspace{\baselineskip}
%
\section{Particle in the field of nonlinear cosmic string}

In accord with the paper [16] consider
motion of nonrelativistic particle in the gravitational field (23).
Corresponding equations of motion
\begin{equation}
\frac{d^2x^k}{dt^2} = - \frac{1}{2}h_{00,k} = \gamma\mu\varepsilon^2
\frac{\eta^k}{r}
\end{equation}
can be easily solved in $\{x,y\}$ plain.
Introducing cylindrical coordinates we have, in the particular case
of rotational motion $(r = R)$, the following equation for the
angular velocity of the particle:
\begin{equation}
\Omega = \varepsilon\frac{\sqrt{\gamma\mu}}{R}.
\end{equation}
Knowing this velocity one can calculate the power of the
corresponding gravitational radiation. Denoting the mass of
particle by $m$ and using results of paper [17] we obtain the power
of radiation:
\begin{equation}
N = \frac{8}{5}c^{-5}m^2l^4\Omega^6
= \frac{8}{5}c^{-5}R^{-2}\gamma^4m^2\mu^3\varepsilon^6.
\end{equation}

Let the mass be equal to mass of proton,
$m = m_p \sim 1,7 10^{-27}kg.$
Thickness of the string, for example, in theories of great
unification, is of the order of the corresponding
Compton wavelength, i.e.
$R \sim 10^{-30}m$ [18] while linear mass density
$\mu \sim 10^{21}kg/m$.
Therefore, power of gravitational radiation per proton
is of the order of
$\sim \varepsilon^6 10^{-14} Watts$.

As to frequency of gravitational radiation, it is two times of
the frequency of orbital motion of the particle, due to [17].
In our case, we thus have
\begin{equation}
f = 2\Omega = 2\varepsilon\frac{\sqrt{\gamma\mu}}{R}.
\vspace{\baselineskip}
\end{equation}
Using the abovementioned numerical values it is easy to find that
the frequency is of the order of
$\sim \varepsilon 10^{27}sec^{-1}$.
We see from this that despite smallness of the parameter
$\varepsilon$ the frequency of gravitational radiation
can be very large (see {\bf Conclusions}).
\vspace{\baselineskip}
%
\section{Equations of motion of nonlinear cosmic string}

Let us find equations of motion of cosmic string taking into account
nonlinear terms as well.
To this end, we insert (15) into energy-momentum conservation law
and use the medium continuousness condition
$\nabla_{\beta}(\mu u^{\beta}) = 0$ in it, and
continuousness condition of each string
$\nabla_{\beta}(\mu l^{\beta}) = 0$.
Then we obtain the following equation:
\begin{equation}
\frac{Du^{\alpha}}{d\tau}
- \frac{Dl^{\alpha}}{d\rho}(1 - \frac{\varepsilon^2}
{2}l^{\gamma}l_{\gamma})
+ \varepsilon^2\frac{Dl^{\gamma}}{d\rho}l^{\alpha}
l_{\gamma} = 0.
\end{equation}
Since spacelike vector
$l^{\alpha}$ in isometric frame of coordinates
$\{\tau,\rho\}$ obey the condition
$\displaystyle\frac{Dl^{\gamma}}{d\rho}l_{\gamma} = 0$[19]
the equations of motion can be written in the form
\begin{equation}
\frac{Du^{\alpha}}{d\tau}
- \frac{Dl^{\alpha}}{d\rho}(1 - \frac{\varepsilon^2}
{2}l^{\gamma}l_{\gamma}) = 0.
\end{equation}
In pseudo-Euclidean space-time it takes simple form
\begin{equation}
\frac{d^2x^{\alpha}}{d\tau^2}
- \frac{d^2x^{\alpha}}{d\rho^2}(1 -
\frac{\varepsilon^2}{2}
\frac{dx^{\gamma}}{d\rho}\frac{dx_{\gamma}}{d\rho}) = 0.
\end{equation}
In Newtonian limit eq.(29) takes the form
\begin{equation}
\frac{d^2x^k}{dt^2}
- \frac{d^2x^k}{d\rho^2}(1 + \frac{\varepsilon^2}{2}
\frac{dx^m}{d\rho}\frac{dx_m}{d\rho}) = 0,
\end{equation}
that completely coincides with wellknown equation describing
nonlinear oscillations of string in classical mechanics
(see, for example ref.[20]).
One can easily prove this taking solution of eq. (31) in the form
$x^k = a^k(t)\sin n\displaystyle\frac{\pi\rho}{L}$.
\vspace{\baselineskip}
%
\section
{Motion of nonlinear cosmic string in (1+1) - dimensional universe}

Let us solve equations of motion (29) in the given gravitational
field. As an example of simple external $V_4$ we take
(1+1)-dimensional Freedman-type space-time, namely,
\begin{equation}
ds^2 = d{x^0}^2 - R(x^0)^2d{x^1}^2 = d{\tau}^2 - R(\tau)^2d{x^1}^2.
\end{equation}
Due to smallness of the coefficient at the nonlinear term, solution
of the eqs.(29) can be presented in series,
\begin{equation}
x^1 = x_0^1 + {\xi}^1,
\end{equation}
and the contribution ${\xi}^1$ have the order of $\varepsilon^2$.
Then, the original equation of motion can be presented as two
equations,
\begin{equation}
\frac{D^2x_0^1}{d{\tau}^2} - \frac{D^2x_0^1}{d{\rho}^2} = 0
\end{equation}
and
\begin{equation}
\frac{d^2\xi^1}{d\tau^2} - \frac{d^2\xi^1}{d\rho^2}
+ \frac{\varepsilon^2}{2}
\frac{d^2x_0^1}{d\rho^2}\frac{dx_0^1}{d\rho}\frac{dx_0^1}{d\rho}.
\end{equation}
General solution of the eq.(34) in the metrics (32) has been
obtained in ref.[21]. Here, we present only the following anzatz:
\begin{equation}
x_0^1 = \Phi(\rho + \tau) - \int\frac{d\tau}{R(\tau)} + C,
\end{equation}
where $\Phi$ is an arbitrary function, $C$ is a constant.
Take this function as propagating sine wave,
$\Phi(\rho + \tau) = a\sin\Omega(\rho + \tau)$.
Inserting the explicit form of propagating wave into eq. (36)
we cast equation of perturbed motion into standard form,
\begin{equation}
\frac{d^2\xi^1}{d\tau^2} - \frac{d^2\xi^1}{d\rho^2}
= \varepsilon^2\Psi(\rho + \tau)
\end{equation}
with the r.h.s.
\begin{equation}
\Psi(\rho + \tau) = {(\frac{A}{2})}^3\Omega^4[\sin3\Omega(\rho
+ \tau) + \sin\Omega(\rho + \tau)].
\end{equation}
Particular solution of this equation can be written
\begin{equation}
\xi^1(\rho,\tau) = \frac{\varepsilon^2}{2}
\int_{0}^{\tau}(\int_{\rho - \tau +
\Theta}^{\rho + \tau - \Theta}\Psi(\chi + \Theta)d\chi)d\Theta.
\end{equation}
Calculation of this integral shows that perturbation function
contains non-periodical terms. In the other words, nonlinearity
of the equations of motion of cosmic string yields changes in the
oscillations amplitude of wave propagating along it, namely, the
amplitude increases in time proportionally to $x^0$.

On the other hand, if we insert the term arising from
multiplication of $\varepsilon^2$ to Riemann-Cristoffel symbols
into (35), that corresponds to the metrics (32), the amplitude
increases in time even more, namely, proportionally to $(x^0)^2$.
\vspace{\baselineskip}
\section{Conclusions}

Deriving nonlinear tension tensor of stringful matter (14)
we obtained space-time interval (23) generated by nonlinear
cosmic string.
We derived also nonlinear equations of motion of cosmic string (29)
containing term of second order in $l^{\alpha}$.
On the basis of this equations we briefly analyzed dynamics
of nonlinear cosmic string in (1+1)-dimensional Freedman model of
the universe.

As to the value of the coefficient
$\varepsilon^2$ entering the expressions that we found,
it seems to be less than $\lambda$ in the lagrangian (5).
Since in the framework of inflationary models
$\lambda \simeq 10^{-14} [15]$ then $\varepsilon \geq 10^{-7}$.

As a consequence, frequency of gravitational radiation of particles
rotating around the cosmic string has a minimal order of
$\sim 10^{20} sec^{-1}$, and corresponding field quanta have
a minimal energy $\sim 10^{-4}Gev$.
This value is much bigger than the energy of gravitons radiated,
for example, by oscillations of loops of linear cosmic string [22],
and also of synchrotron radiation arising from kink-like
deformations of linear cosmic string [23].

 The considered above nonlinear string model can also
be applicated as in [19] to the compaund hadron conception,
for the meson describing, in particularly.

 And at last, it is necessary to point out that the discussing
string model is described by such nonlinear hyperbolic equations
that drastically differ from the wellknown Liuwille [19] and
Korteweg - de Vries equations.\\

\section*{Acknowledgments}

This work is partially supported by
{\it Organization and evolution of natural structures} program,
Ministry of Sciences -- Academy of Sciences, Kazakstan.

\newpage
\indent {\bf Fig.1.}\\
{\it Available on request from authors via e-mail .}

\end{document}